# *In-situ* Thermophysical Measurement of Flowing Molten Chloride Salt Using Modulated Photothermal Radiometry

Ka Man Chung[1,#], Ye Zhang[3,#], Jian Zeng[2], Fouad Haddad[3], Sarath Adapa[2], Tianshi Feng[2], Peiwen Li[3], Renkun Chen[1,2,*]

[1]Material Science and Engineering Program, University of California, San Diego, La Jolla, California 92093, United States

[2]Department of Mechanical and Aerospace Engineering, University of California, San Diego, La Jolla, California 92093, United States

[3]Department of Aerospace and Mechanical Engineering, The University of Arizona, Tucson, AZ 85721, United States

* Corresponding author.

Email address: rkchen@ucsd.edu (R. Chen)

# These authors contributed equally

## Abstract

Molten salts are a leading candidate for high-temperature heat transfer fluids (HTFs) for thermal energy storage and conversion systems in concentrated solar power (CSP) and nuclear energy power plants. The ability to probe molten salt thermal transport properties in both stationary and flowing status is important for the evaluation of their heat transfer performance under realistic operational conditions, including the temperature range and potential degradation due to corrosion and contamination. However, accurate thermal transport properties are usually challenging to obtain even for stagnant molten salts due to different sources of errors from convection, radiation, and corrosion, let alone flowing ones. To the best of authors' knowledge, there is no available *in-situ* technique for measuring flowing molten salt thermal conductivity. Here, we report the first *in-situ* flowing molten salt thermal conductivity measurement using modulated photothermal radiometry (MPR). We could successfully perform the first *in-situ* thermal conductivity measurement of flowing molten $NaCl-KCl-MgCl_2$ in the typical operating temperature (520 and 580 °C) with flow velocities ranging from around 0.3 to 1.0 m s$^{-1}$. The relative change of the molten salt thermal conductivity was measured. Gnielinski's correlation was also used to estimate the heat transfer coefficient *h* of the flowing $NaCl-KCl-MgCl_2$ in the given experimental condition. The work showed the potential of the MPR technique serving as an *in-situ* diagnostics tool to evaluate the heat transfer performance of flowing molten salts and other high-temperature HTFs.

1. **Introduction**

Concentrated solar power (CSP) plants with thermal energy storage (TES) systems are one of the most promising emerging renewable energy systems with the capability of delivering dispatchable electricity by utilizing abundant solar energy (Bauer et al., 2012; Palacios et al., 2020). To deliver the heat collected from the concentrated sunlight to a heat exchanger and thermal storage tank, a heat transfer fluid (HTF) is needed for a CSP system. An ideal HTF requires a high thermal conductivity for efficient thermal energy transfer, a high heat capacity for large quantities of sensible thermal energy storage, and a low viscosity that renders minimal power consumption for circulating the fluid (Vignarooban et al., 2015; Weinstein et al., 2015).

In state-of-the-art CSP technologies, there are two common HTFs: oil-based HTFs and salt-based HTFs (molten salts). Synthetic oil, such as Therminol VP-1 (Biphenyl and Diphenyl oxide), is a commonly used HTFs in the 1$^{st}$ and 2$^{nd}$ generation of CSP power plants (Merchán et al., 2022; Vignarooban et al., 2015; Weinstein et al., 2015). It can reach a maximum operating temperature of 400 ºC. However, oil-based CSP plants usually have limited thermal energy storage capability (Merchán et al., 2022; Tian and Zhao, 2013), which makes molten salts being competitive as the HTFs for CSP applications from the energy storage capability point of view. Nitrate salt mixtures, $NaNO_3$-$KNO_3$, is used for heliostat CSP tower with a maximum operating temperature of 560 ºC (Bauer et al., 2013). CSP plants using molten $NaNO_3$-$KNO_3$ can provide around 8-10 hours of thermal energy storage (Tian and Zhao, 2013). The next generation CSP integrated with TES facilities aims at a higher operating temperature (> 700 ºC) to achieve both a high power cycling efficiency (> 50 %) and a lower levelized cost of electricity (LCOE) of $ 0.05 per kWh (Merchán et al., 2022). Eutectic molten chloride salts, such as $MgCl_2$-$KCl$, or $NaCl$-$KCl$-$MgCl_2$, that can work at high temperatures of up to 800 ºC, are considered as good HTF candidates for next-generation CSP due to their favorable thermal and transport properties, high boiling point, acceptable low freezing point, and low price to meet the large quantity demand for thermal storage (Li et al., 2016; Wang et al., 2023; Zhao and Vidal, 2020).

However, limited to reliable molten salt thermal conductivity data, it is usually difficult to evaluate molten salt heat transfer performance in the entire operating temperatures. For example, there is a difference of up to 40% among the reported thermal conductivity values for solar salt (Pfleger et al., 2015), despite its widespread use in commercial CSP plants. Reliable data on emerging high-temperature chloride salts are even more scarce. The scattered thermal conductivity data in the molten salt is mainly due to measurement errors, caused by convection, radiation, and corrosion during the measurement process. These errors can be further magnified in the convectional measurement techniques (i.e., steady state method, transient hot-wire (THW) method, and laser flash analyzer (LFA) at high temperatures. Other than the importance of obtaining accurate molten salt thermal conductivity when it is static, it is more important to monitor the thermal transport properties of molten salts in their flowing state in real-world applications. The effective thermal conductivity $k_{eff}$ of flowing molten salts is a crucial indicator to evaluate their thermal transport property under various flow and temperature operational conditions. Several measurement techniques have been developed and modified to remove the barriers in measuring molten salts, molten metal, and other high-temperature fluids. Zhao *et al.* (Zhao et al., 2020) have developed a frequency-domain THW method to measure molten nitrate salts. Their reported results are believed to be a more appropriate representative of molten nitrate salt thermal conductivity because their frequency-domain method can limit the thermal penetration of the measurements within less than 100 μm, thus enabling a neglectable natural convection error in the measurements. However, their measurements were still limited within the operating temperatures of molten nitrate salts (less than 500 ºC). At even higher temperatures, the alumina-coated Pt wire in their measurement device could be damaged due to thermal stress and corrosion issues at higher temperatures.

*In-situ* measurement of flowing fluids is also important to the evaluation of the thermal transport properties of flowing fluids in real time. However, there is a lack of measurement techniques for flowing fluid. Langstaff *et al.* (Langstaff et al., 2013) have developed an *in-situ* thermophysical measurement technique for high-temperature liquids using aerodynamic levitation (ADL). They demonstrated the technique by measuring the thermophysical property of $Al_2O_3$ droplets *in-situ* over the entire liquid and super-cooled regimes. However, this technique might not be accessible to flowing liquids. Hong and Kim (Hong and Kim, 2012) measured flowing D.I. water in a circular tube using a 3ω method. It demonstrates that the intrinsic thermal conductivity of the fluids can be measured when the flow condition and heating frequency are carefully adjusted. However, the technique may not be suitable for high-temperature measurements where the

3ω heater and sensor can be damaged at high temperatures. Mehrvand and Putnam (Mehrvand and Putnam, 2017) reported the use of time-domain thermoreflectance (TDTR) to characterize the local heat transfer coefficient of flowing water and refrigerant in a microchannel. The difficulty of the method may have arisen from the preparation complexity and the stability of the metal coating on the surface of the measurement window. In addition, further work is needed to demonstrate the capability of the technique to perform measurements for wider channels and at high temperatures (> 500 ºC) for more practical applications in industry and power plants where HTFs are usually operating at high temperatures.

Recently, we have reported the thermal conductivity measurement of molten salts, including molten nitrate and chloride salts, using the modulated photothermal radiometry (MPR) technique. It is a laser-based, non-contact, frequency-domain measurement technique that can eliminate the errors from corrosion and convection effects in molten salts, thus being capable of extracting accurate molten salt thermal conductivity. The MPR technique was also applied for measuring flowing fluids *in-situ* at room temperature and intermediate temperatures (Zeng et al., 2021a). In the MPR flowing fluid property measurement, the intrinsic thermal conductivity $k_0$ of D.I. water and thermal oil were extracted when the flows satisfy the criteria of $Re^{1/2}Pr^{1/3} < 100$ in the given experimental conditions. In this work, we performed the *in-situ* thermal conductivity of flowing molten salt inside a molten salt flow loop. We selected NaCl-KCl-MgCl$_2$ (wt% 15.11–38.91–45.98) to study the relative changes in $k_{eff}$ of the molten salt under different flow conditions and temperatures. Based on the measured intrinsic thermal conductivity, the convective heat transfer coefficient $h$ of flowing NaCl-KCl-MgCl$_2$ was also estimated using Gnielinski's correlation.

To our best knowledge, it is the first *in-situ* thermal conductivity measurement of flowing molten salt. It opens the possibility to apply the MPR technique to accurately extract and monitor possible changes in the thermal transport properties of molten salts and other high-temperature liquids (i.e., molten metal) under various operating conditions, which will be valuable to the further optimization of the overall heat transfer performance of the HTFs.

## 2. Experimental Procedures

### 2.1. Salt Properties and Salt Preparation

The molten salt loop equipped with the molten salt reservoir was built at the University of Arizona (UA). The selected chloride salt in this work is NaCl-KCl-MgCl$_2$ (wt% 15.11–38.91–45.98) (Wang et al., 2021; Zhao and Vidal, 2020). The overview of the molten salt loop integrated with the MPR setup is shown in Figure S1. The physical properties of the molten salt studied by the team at the UA and National Renewable Energy Laboratory (NREL) (Wang et al., 2021) are listed in **Table 1**.

The constituent components of the eutectic salt, including anhydrous salt powders of MgCl$_2$ (Thermo Fisher Scientific Chemical Inc., purity: 99.0 %), KCl (Thermo Fisher Scientific Chemical Inc., purity: 99.0-100.5 %), and NaCl (Thermo Fisher Scientific Chemical Inc., purity: 99.5 %) are commercially available. They were weighted according to the required weighted percentage of the composition (45.98% MgCl$_2$, 38.91% KCl, 15.11% NaCl). Less than 0.1 wt % of elemental Mg was added as suggested by Zhao and Vidal (Zhao and Vidal, 2020) to reduce the presence of corrosive impurities, mainly water. The weighted salts were first ground and mixed in a tumbler mixer for more than 200 hours to ensure that the components turned into fine powders and mixed well. The eutectic mixtures were then loaded into a Hastelloy C-276-made reservoir (inner diameter: 508 mm, height 533.4 mm) for melting. The schematic of the reservoir is shown in **Figure 1**. The reservoir was sat in a furnace (Model KM714, Skutt Ceramic Products, Inc.) that can heat the molten salt from room temperature to 750°C. Before the salt melting process, the reservoir was evacuated to remove air, moisture, and other impurities. The salt was first heated to 150 °C and stayed at that temperature for more than an hour. It is for the purpose of removing the water content in the salt. After the salt was melted, dry nitrogen gas was purged into the reservoir to further remove the residual water and air that were potentially trapped inside the salt. Humidity and oxygen sensors were installed at the opening of the vent of the reservoir for air and moisture monitoring. The volume of the molten salt prepared in this work was about 102.68 L.

## 2.2. High-Temperature Molten Salt Loop and Operation Procedures

**Figure 2 a** shows the schematic of the molten salt loop and the *in-situ* MPR measurements. The molten salt loop was mainly comprised of a molten salt reservoir, a high-temperature molten salt pump (Nagle Pumps, Inc.) with a maximum molten salt pumping rate of 0.63 L s$^{-1}$, and a designated MPR test section. The materials for the flow pipes of the loop were made of Hastelloy C-276 (outer diameter: 25.4 mm, wall thickness: 1.2446 mm; total length: 6.4 m). The flow loop in total could contain a molten salt volume of 3.24 L. The MPR test section was a modified Inconel 625 tube with a tube outer diameter of 25.4 mm and 1 mm in wall thickness. In the middle of the tube, which is the MPR measurement window, a 100 μm thick Inconel 625 sheet was wielded on the tube. Such a thin layer is required to maintain the high sensitivity of the measurement to the flowing molten salt regime. The thin Inconel 625 sheet was coated with a thin layer of Pyromark 2500 (10-15 μm) for laser heating absorption. The back side of the Inconel 625 sheet was also coated with a 200 nm thick Pt layer to avoid severe corrosion of the measurement window. The photograph of the MPR test section is given in Figure S2. The pipes of the flowing loops were tilted by 5 ° relative to the ground so that the molten salt could flow back to the reservoir in case the molten salt flow loop stopped operating.

After being pumped out, the molten salt flows into the test section for the MPR *in-situ* thermal conductivity measurement. The molten salt flow velocity was measured using an ultrasonic flow meter (FLUXUS® ADM 7407, Flexim Americas Corp.). Ceramic tubular heaters with controlled power (VF-180-3-19.75-V-S, Thermcraft Inc.) and extra heating tapes (STH051-080, Omega Engineering, Inc.) were also used to keep all the pipes at a temperature above 469 °C for the startup of the operation of the loop. Benchtop On/Off digital temperature controllers (SDC240KC-A, Semi-conductor Smart Solutions, LLC.) were used to control the temperature of the pipes beyond the test section. The power supplied to the heaters and the temperature of different sections of the molten salt flow loops were monitored by a power controller (uF1HXTA0-16-P1R00, Control Concepts Inc.) and the thermocouples coupled with a multifunction I/O devices (NI cDAQ-9178, NI9229, National Instruments.) respectively. The specification of the employed instruments including measurable parameters, accuracy, model and maker, and measurable range are listed in Table S1.

## 2.3. MPR *in-situ* Measurement Principle

### 2.3.1. MPR Model for Flowing Fluid Measurement

In the MPR measurement, a continuous laser with modulating angular frequency $\omega$ ($q(t) = q_s e^{j\omega t}$) heats the front surface of the Inconel window of the test section. The temperature oscillation of the fluid flowing inside the test section ($\theta_s(t) = \theta_s e^{j\omega t}$) is measured using both a high-speed MCT (Mercury-Cadmium-Telluride) infrared (IR) detector and a pyrometer (Lumasense Technologies IGA 320/23- LO). The MCR detector is connected to a lock-in amplifier for collecting the harmonic voltage signal induced by the oscillating thermal emission in the entire frequency range. The pyrometer, on the other hand, records the actual $|\theta_s|$. The obtained voltage signal is converted to $|\theta_s|$ based on the $|\theta_s|$ vs. voltage calibration curve. The instrumentation details of the MPR system are described in our earlier work (Zeng et al., 2021b).

The surface temperature oscillation on the cylindrical (curved) surface induced by an intensity-modulated laser with an angular frequency $\omega$ is given by (Wang et al., 2004):

$$\theta_{s,cyl}(r,\phi,\omega) = \left( \frac{q_s}{2\pi k_s} \frac{2I_0(\sigma r)}{I'_0\left(\frac{\sigma D}{2}\right)} \sin\left(\frac{\theta_0}{2}\right) + \frac{I_1(\sigma r)}{I'_1\left(\frac{\sigma D}{2}\right)} (\theta_0 + \sin(\theta_0)) \times \cos\left(\frac{\pi}{2} - \phi\right) + 2\sum_{m=2}^{\infty} \frac{I_m(\sigma r)}{I'_m\left(\frac{\sigma D}{2}\right)} \times \cos\left[\frac{m}{2}(\pi - 2\phi)\right] \times \left[ \frac{\sin\left(\frac{(m+1)\theta_0}{2}\right)}{m+1} + \frac{\sin\left(\frac{(m+1)\theta_0}{2}\right)}{m-1} \right] \right) \quad (1)$$

where $k_s$ is the thermal conductivity of the cylindrical material, $\theta_0$ is the subtended angle by the laser spot, and $\phi$ is the azimuth angle; $I_m$ is the $m^{\text{th}}$ order of the modified Bessel function of the first kind and $I'_m$ is the derivative of $I_m$; and $\sigma = \sqrt{\frac{i\omega}{\alpha_s}}$, where $\alpha_s$ is the thermal diffusivity of the cylindrical material. Wang *et al.*'s numerical calculation (Wang et al., 2004) shows that when the diameter of the cylindrical tube increases and the thermal penetration depth $L_p$ is much shorter than the laser spot size, the solution given in Eq. (1) is reduced to the case for a planar plate under a modulated laser heating:

$$\theta_{s,pl}(0,\omega) = q_s \left( \frac{\exp\left(-\frac{\pi}{4}j\right)}{e_s\sqrt{\omega}} \right) \quad (2)$$

where $e_s$ is the thermal effusivity of the plate. From our previous analysis of the MPR measurement of flowing fluids inside a cylindrical tube (Zeng et al., 2021a), it has been shown that the percentage difference $\theta_{s,cyl}$ and $\theta_{s,pl}$ is up most 7% under our given experimental conditions where the modulated frequency range from 0.5 to 2 Hz, the tube diameter $D = 25.4$ mm, and the laser spot size $D_{laser} = 10$ mm. The analysis provides a reasonable basis for treating MPR measurements of flowing fluid confined by a cylindrical tube as a simple planar geometric heat transfer problem. In the following analysis, we would simplify the problems by considering Eq. (2) to avoid the mathematical complexity.

The flowing molten salt inside the test section, the 100 μm Inconel 625 window on the test section, and the Pyromark coating, together form a three-layer structure. Since the laser heating spot size (~10 mm) is much larger than the thermal penetration depth $L_p$ (< 0.5 mm) within the modulation frequency used in this study, we can assume that the heat transfer is one-dimensional (1D). The 1D heat transfer model for a multi-layer sample is in the form of (Zeng et al., 2021a; Zeng et al., 2021b),

$$\begin{bmatrix} \theta_b \\ q_b \end{bmatrix} = M_n M_{n-1} \cdots M_i \cdots M_1 \begin{bmatrix} \theta_s \\ q_s \end{bmatrix} = \begin{bmatrix} a(i\omega) & b(i\omega) \\ c(i\omega) & d(i\omega) \end{bmatrix} \begin{bmatrix} \theta_s \\ q_s \end{bmatrix} \tag{3}$$

where $\theta_s$ and $\theta_b$ are the temperature oscillation on the top layer and the bottom layer, respectively; $q_s$ and $q_b$ are heat flux imposed on the top layer and the bottom layer, respectively.

The transfer matrix $M_i$ is given by:

$$M_i = \begin{bmatrix} \cosh(D_i\sqrt{j\omega}) & -\dfrac{\sinh(D_i\sqrt{j\omega})}{e_i\sqrt{j\omega}} \\ -e_i\sqrt{j\omega}\sinh(D_i\sqrt{j\omega}) & \cosh(D_i\sqrt{j\omega}) \end{bmatrix} \tag{4}$$

where $D_i = \dfrac{l_i}{\sqrt{\alpha_i}}$. $l_i$ and $\alpha_i$ are the thickness and the thermal diffusivity of the $i$ th layer respectively; $e_i = \sqrt{\rho_i c_{pi} k_i}$ is the thermal effusivity of the $i$ th layer which is a function of the density $\rho_i$, specific heat capacity $c_{pi}$, and thermal conductivity $k_i$ of the $i$ th layer, respectively. When $L_p$ is long enough to probe the molten salt layer, the surface temperature oscillation in a simplified

expression in the low-frequency-limit, is reduced into the following: (Zeng et al., 2021b; Zhao et al., 2016)

$$\theta_s = q_s \left( \frac{\exp\left(-\frac{\pi}{4}j\right)}{e_o \sqrt{\omega}} + R_{c,s} \right) \tag{5}$$

where $e_o$ is the thermal effusivity of molten salt; $R_{c,s}$ is the lumped thermal resistance from the Inconel sheet wielded on the MPR test section and the Pyromark coating:

$$R_{c,s} = \left(1 - \frac{e_s}{e_o}\right) \frac{l_s}{k_s} + \left(\frac{1}{2} - \frac{e_c^2}{e_o^2}\right) \frac{l_c}{k_c} \tag{6}$$

where $e$, $k$ and $l$ are thermal effusivity, thermal conductivity, and thickness. The subscripts $s$ and $c$ stand for the thin Inconel layer and the coating layer. According to Eq. (5), the thermal effusivity of the molten salt layer $e_o$ can then be derived from the slope $m_o$ of the $|\theta_s|$ vs. $\omega^{-\frac{1}{2}}$ plot in the low-frequency region given that $q_s$ is known. To eliminate the uncertainty in $q_s$ measurement, a reference is introduced for calibrating $q_s$. Using a differential method, $e_o$ can be obtained by comparing $m_o$ and $m_{ref}$, the slope of the $|\theta_s|$ vs. $\omega^{-\frac{1}{2}}$ plot of the reference with a standard thermal effusivity ($e_{ref}$): $e_o$ can be obtained as given in Eq. (7), by comparing with the reference.

$$e_o = \frac{m_{ref}}{m_o} e_{ref} \tag{7}$$

Borosilicate glass was chosen as the reference sample because it has a well-established thermophysical property (Assael et al., 2004).

$e_o$ of the molten salt layer is then converted to $k_o$,

$$k_o = \frac{e_o^2}{\rho_o c_{p,o}} \tag{8}$$

The thermal conductivity calculation based on Eq. (7-8) takes advantage of the simplicity of mathematical operation without involving complicated model fitting and data processing. It provides a simple and convenient method to dictate the sample's thermal conductivity by comparing the slope of $|\theta_s|$ vs. $\omega^{-\frac{1}{2}}$ plot in the appropriate frequency range. It benefits *in-situ* measurements that require a timely response to the change in thermal conductivity of the layer that is of interest. The limitations of this method will be discussed in *Section 3.2*.

### 2.3.2. Criteria of Measuring Intrinsic Thermal Conductivity of Flowing Fluids

**Figure 2 b** shows the 2D schematic of the flow inside the test section during the MPR measurement. The thermal boundary layer, momentum boundary layer, and the thermal penetration depth at both high-and-low frequency cases are illustrated. For convenience in the flow analysis, we assume a laminar flow. The hydrodynamic entrance length $L_{ef}$ and the thermal entrance length $L_{eh}$ are given in the following (Ting, 2022):

$$L_{ef} = 0.05 Re_D D \tag{9}$$

and

$$L_{eh} = 0.05 Re_D Pr D \tag{10}$$

where $Re_D$ is the Reynold number, $Pr$ is the Prandtl number, and $D$ is the cylindrical tube diameter. Based on the thermophysical property of molten NaCl-KCl-MgCl$_2$ and the experimental conditions given in Table 1, $Re_D$ and $Pr$ of the molten salt ranges from around 3805 – 15000 and 7, respectively. Using Eq. (9), the estimated $L_{ef}$ and $L_{eh}$ ranges from 500 – 1900 cm, and 3400 – 13000 cm, respectively. Since the laser spot of ~ 10 mm targets at the middle of the MPR test section where the Inconel measurement window locates, the flow is still in both hydrodynamic and thermal entrance regions. The above analysis shows that the thermal boundary layer δ of the flow is always thinner than its momentum boundary layer Δ in our measurements. According to our previous MPR flowing fluid measurement (Zeng et al., 2021a), when the thermal penetration depth in the measurement is less than both the thermal boundary layer and momentum boundary layer, the intrinsic thermal conductivity of a flowing fluid can be measured (**Figure 2 b upper schematic**). In other words, the following criteria has to be satisfied: $L_p$ < min [ Δ, δ]. Otherwise, when $L_p$ > Δ or δ, the temperature response from the test section would be affected by the momentum boundary layer, a more prominent convective effect would then be observed. Under the laminar flow assumption, $\delta = Pr^{-\frac{1}{3}}\Delta$, and $\Delta = 5\sqrt{\frac{vx}{u_b}}$, where $v$ is the kinematic viscosity of the fluid, and $x$ is the characteristic entrance length of the measurement section as shown in **Figure 2 b** ($x \sim$ 15 cm in our experiment), and $u_b$ is the flow velocity. To satisfy the criteria, we require $L_p < 0.2\delta$. The criterion can then be expressed with the flowing relation (Zeng et al., 2021a):

$$Re_D^{\frac{1}{2}}Pr^{-\frac{1}{3}} < \sqrt{\frac{DD_{laser}\pi f}{\alpha_o}} \quad (11)$$

Given that $D$ = 25.4 mm and $D_{laser}$ = 10 mm in our experimental work, and using $\alpha_o$ of the molten salt (Wang et al., 2021), when $f$ is within 0.5 to 2 Hz, $Re_D^{\frac{1}{2}}Pr^{-\frac{1}{3}} <$ 100, the forced convection effect is neglected. When $Re_D^{\frac{1}{2}}Pr^{-\frac{1}{3}} \gg$ 100, the forced convection effect is no longer negligible. $k_{eff}$ that is higher than the $k_o$ of the flow fluid is thus obtained in the measurement. For convenience, we define $\lambda = \frac{k_{eff}}{k_o}$ as the normalized thermal conductivity of the flowing fluids. It is an indicator that describes the relative change in the thermal transport property of the flowing fluids due to the convection effect.

## 3. Results and discussions

### 3.1. MPR measurement results

**Table 2** summarizes the experimental conditions of the MPR flowing molten salt measurements. To illustrate the expected experimental results, we built a COMSOL model using a laminar flow module coupled with heat transfer to study the thermal response of flowing molten salt inside the MPR test section. The test section and material properties NaCl-KCl-MgCl$_2$ at $T$ = 520 °C are from reference (Wang et al., 2021). **Figure 3a** shows the COMSOL simulation result of the MPR flowing molten salts NaCl-KCl-MgCl$_2$ measurement at $T$ = 520 °C in the entire frequency range with $u_b$ changes from 100 to 1000 mm s$^{-1}$. From the COMSOL simulated results of $|\theta_s|$ vs. $\omega^{-\frac{1}{2}}$, three regions are observed: the flowing molten salt region in the low-frequency range, the Inconel alloy region in the intermediate-frequency range, and the Pyromark coating region in the high-frequency region. The slopes of the $|\theta_s|$ vs. $\omega^{-\frac{1}{2}}$ plots in the low-frequency region decrease gradually with the increasing $u_b$. When $u_b$ > 500 mm s$^{-1}$, the slope starts to plateau. It is a result of the forced convection effect, equivalently higher $k_{eff}$ at high flow velocity, leading to a less prominent $|\theta_s|$. **Figure 3 b** shows experimental $|\theta_s|$ vs. $\omega^{-\frac{1}{2}}$ plot of molten NaCl-MgCl$_2$-KCl at $T$ = 520 °C with different $u_b$ of 0.375, 0.658, and 1.000 m s$^{-1}$. Despite the discrepancy between our experimental results and the COMSOL simulation results, we could observe the change in the slope in the low-frequency region. It is consistent with the COMSOL simulation results. The $|\theta_s|$ vs. $\omega^{-\frac{1}{2}}$ curves show the transition from the molten salt region to the Pyromark-coated Inconel sheet region when the modulation frequency increases. The slopes of the $|\theta_s|$ vs. $\omega^{-\frac{1}{2}}$ plots in the low-frequency region decrease along with increasing $u_b$. At $u_b$ = 1.0 m s$^{-1}$, the slope nearly plateaus. It indicates the significant convection effect at high flow velocities, as the model predicts. Using Eqs. (7) and (8), $k_{eff}$ of the flowing molten salts was obtained. The $|\theta_s|$ vs. $\omega^{-\frac{1}{2}}$ data of the flowing molten salt in the low-frequency range (0.5 – 1.5 Hz) at $T$ = 520 °C and $T$ = 580 °C with different $u_b$ ranging from ~ 0.27 to 1.04 m s$^{-1}$ are included in the Figure S3. **Figure 4** shows the normalized thermal conductivity $\lambda$ as a function of $u_b$. $\lambda$ approaches unity when the flowing molten salts exhibit $k_o$ (Wang et al., 2021); otherwise, it increases as a function of $T$ and $u_b$. From the experimental results, $\lambda$ of flowing molten salt is a strong function of $u_b$. When $u_b$ is small ($u_b$ < 0.4 m s$^{-1}$), we obtained $\lambda$ close to 1. At $T$ = 520 °C, the measured $k_{eff}$ of molten NaCl-MgCl$_2$-KCl with $u_b$ = 0.375 m s$^{-1}$ is 0.477 W m$^{-1}$ K$^{-1}$; at $T$ = 580 °C, the measured $k_{eff}$ of the molten salt $u_b$ = 0.266 m s$^{-1}$ is 0.424 W m$^{-1}$ K$^{-1}$. Upon $u_b$ increases, $\lambda$ increases substantially. At $T$ = 520 °C with $u_b$ = 1.0 m s$^{-1}$, $\kappa$ increases to 6.40; at $T$ = 580 °C with $u_b$ = 1.05 m s$^{-1}$, $\lambda$ increases to 7.44. The enhancement in the $k_{eff}$ is due to the forced convection effect in the flowing molten salt. The $k_{eff}$ enhancement of flowing molten salt NaCl-MgCl$_2$-KCl behaviors similarly at both $T$ = 520 °C and 580 °C. However, the $k_{eff}$ enhancement at $T$ = 580 °C begins at a slightly lower $u_b$. It implies that the forced convection effect is more significant at higher temperatures, due to a lower viscosity of the molten salt.

The experimental results are further analyzed by plotting $\lambda$ as a function of $Re_D$ and $Re_D^{\frac{1}{2}} Pr^{-\frac{1}{3}}$ respectively, as shown in **Figure 5 a** and **Figure 5 b**. On the $\lambda$ vs. $Re_D$ plot shown in **Figure 5 a**, $\lambda$ increases monotonically as a function of $Re_D$. For flowing fluid in a smooth tube, the flow transits to turbulent when $Re_D$ ~ 4000 and the turbulence is fully developed when $Re_D$ > 10,000 (Mills and Coimbra, 2015). It explains the enhancement of $\lambda$ at large $Re_D$ values and it is consistent with the results shown in **Figure 4**. **Figure 5 b** shows that when the flowing molten salt has relatively small $Re_D^{\frac{1}{2}} Pr^{-\frac{1}{3}}$ values, the measured $\lambda$ converges to unity; when $Re_D^{\frac{1}{2}} Pr^{-\frac{1}{3}}$ increases and much beyond 100, the measured $\lambda$ increases significantly. It is consistent with our analysis of the criteria of intrinsic thermal conductivity measurement of flowing fluids. Our previous flowing D.I. water and thermal oil results are also plotted in **Figure 5 b** for comparison.

It is rarely peculiar that our experimental results shown in the $|\theta_s|$ *vs.* $\omega^{-\frac{1}{2}}$ plots (**Figure 3 b**) exhibit different trends, especially in the intermediate-to-high frequency range, compared to COMSOL model prediction. The reason is still not clear. The scale of the molten salt loop that we used in this work is much larger than the setup shown in other similar studies (Arora et al., 2021; Gill et al., 2013; Sabharwall et al., 2010; Zhang et al., 2022), thus, complicating the measurements and the stability of the molten salt flow. On the other hand, corrosion of the measurement window could also cause the discrepancy. As we mentioned in *Section 2.2*, a thin layer of Pt was coated on the back side of the 100 μm Inconel 625 measurement MPR window on the test section to prevent corrosion. However, the corrosion of the thin Inconel sheet was still inevitable. According to the studies of the corrosion of Inconel alloy under molten chloride salts (Sun et al., 2018), corrosion would make the alloy thinner due to the depletion of individual metal components. On the interface of the corroded alloy, there is a formation of a metal oxide layer, which could show different thermophysical properties compared to the original alloy. The corrosion issue could be attributed to the discrepancies in the intermediate-to-high-frequency range where the thermal response of the test section is more dominant by the thermophysical properties of the alloy and the coating. Nevertheless, according to Eq. (6), the low-frequency thermal response is still more determined by the thermophysical property of molten salt. This explains that our molten salt thermal conductivity calculation is less affected by the effects of molten salt corrosion of the measurement window.

## 3.2. Discussion on the $k_{eff}$ calculation Using the Differential Method

Based on Eqs. (5-8), the thermal conductivity of fluids can be obtained using a differential method. The simplicity of the method allows us to measure $k_o$ without exactly knowing the required physical quantities (e.g., heat flux, and laser spot size). However, one has to be cautious about applying this method in three-layer measurements. The simplification shown in Eq. (5) is only valid if the intermediate layer exhibits a negligible thermal resistance (i.e., negligible thickness). The intermediate layer contributes a heat spreading effect, thus, affecting the heat transfer from the top layer (Pyromark coating) to the bottom layer (molten salt layer). The calculation of $k_o$ would lead to a certain deviation if the differential method is applied. To illustrate this, we compare the simulation $|\theta_s|$ vs. $\omega^{-\frac{1}{2}}$ curves of a three-layer structure (Pyromark coating, 100 μm Inconel alloy, and molten salt layer) and that of a two-layer structure (Pyromark coating, and molten salt layer). It is included in Figure S4. The slopes in the low-frequency region of the $|\theta_s|$ vs. $\omega^{-\frac{1}{2}}$ show a deviation of around 7 %. The deviation can be minimized if the thickness of the Inconel sheet is further reduced.

In our case, the intermediate layer in our study is a 100 μm Inconel sheet. The choice of the Inconel sheet thickness is a compromise between the mechanical strength of the Inconel sheet to contain the flowing molten salt inside the test section and the sensitivity of the amplitude signal ($|\theta_s|$) in the low-frequency region. Nevertheless, in the differential method, we can indicate the relative change of $k_{eff}$ of flowing molten salt by dictating the change in the low-frequency slopes in the $|\theta_s|$ vs. $\omega^{-\frac{1}{2}}$ plots.

## 3.3. Heat Transfer Coefficient Prediction using Gnielinski's Correlation

It is of interest to study the convective heat transfer coefficient $h$ of flowing NaCl-KCl-MgCl$_2$ inside the smooth pipes. Zhang *et al.* (Zhang et al., 2022) recommend that Gnielinski's Correlation is a better correlation for current molten chloride salt systems. Gnielinski's correlation is given by (Mills and Coimbra, 2015):

$$Nu = \frac{(\frac{f}{8})(Re-1000)Pr}{1+12.7(\frac{f}{8})^{0.5}\left(Pr^{\frac{2}{3}}-1\right)} \tag{12}$$

where $Nu = \frac{hD}{k_o}$ is the Nusselt number; $f$ is the Darcy friction factor (Mills and Coimbra, 2015):

$$f = (0.79 \, \text{Ln}(Re_D) - 1.64)^{-2} \tag{13}$$

Gnielinski's correlation is valid when $0.5 < Pr < 2000$ and $3000 < Re_D < 5 \times 10^6$. Using Eqs. (12-13), we calculate $Nu$ and $h$ values the flowing molten salt. The measured intrinsic thermal conductivity of the flowing molten salt ($k_o$ = 0.477 W m$^{-1}$ K$^{-1}$ when $T$ = 520 °C and $u_b$ = 375 mm s$^{-1}$; $k_o$ = 0.424 W m$^{-1}$ K$^{-1}$ when $T$ = 580 °C and $u_b$ = 266 mm s$^{-1}$), the flow velocity in our study, and the known thermophysical properties of the molten salt from reference (Wang et al., 2021) are the input parameters. **Figure 6 a** and **Figure 6 b** show the $Nu$ vs. $Re_D$, and $h$ vs. $u_b$ plots at $T$ = 520 and 580 °C, respectively. The measured values of $Nu$ and $h$ of molten NaCl-KCl-ZnCl$_2$, another eutectic chloride salt with similar compositions, are also plotted alongside for comparison (Zhang et al., 2022). The predicted $h$ of molten NaCl-KCl-MgCl$_2$ increases linearly with $u_b$. It is the result of the turbulence of the flow at high flow velocities. The predicted $h$ of molten NaCl-KCl-MgCl$_2$ shows around 20 % higher than the measured $h$ of molten NaCl-KCl-ZnCl$_2$ in the entire flow velocity range. The discrepancy can be attributed to the limited prediction ability of the correction and the difference in the intrinsic thermophysical properties of these two molten eutectic chloride salts (Li et al., 2016; Wang et al., 2021). Yet, the predicted $h$ of molten NaCl-KCl-MgCl$_2$ using Gnielinski's correlation can provide a very valuable reference and evaluation of the heat transfer performance of molten chloride salt for CSP applications.

## 4. Conclusion

This paper reports the first *in-situ* thermal conductivity measurement of flowing molten salt (NaCl-KCl-MgCl$_2$) from 520 and 580 °C using the MPR method with flow velocities ranging from around 0.3 to 1.0 m s$^{-1}$. The intrinsic thermal conductivity of the molten salt flow can be obtained when the flow velocity is low and when the criterion $Re_D^{\frac{1}{2}}Pr^{-\frac{1}{3}} < \sqrt{\frac{DD_{laser}\pi f}{\alpha_o}}$ is satisfied in the measurement. In our experimental setting, the criterion is given by $Re_D^{\frac{1}{2}}Pr^{-\frac{1}{3}} < 100$. It is consistent with our previous MPR flowing fluid measurements of D.I. water and thermal oil. We were able to detect the relative change of $k_{eff}$ as the turbulence of the molten salt flow developed. Not only being able to measure the intrinsic thermal conductivity of flowing molten salt and the relative change in the effective thermal conductivity under controlled flow conditions but also the heat transfer coefficient of the molten salt can also be obtained using the MPR method. The heat transfer coefficient of molten NaCl-KCl-MgCl$_2$ was obtained using Gnielinski's correlation together with the measured intrinsic thermal conductivity. The heat transfer coefficient of molten NaCl-KCl-MgCl$_2$ obtained in our measurement shows good agreement with the experimentally determined $h$ of another eutectic chloride salt with similar compositions. Our heat transfer coefficient calculation of molten NaCl-KCl-MgCl$_2$ provides a useful reference and evaluation of molten chloride salt heat transfer performance for CSP applications. The work demonstrates that the MPR can serve as an *in-situ* diagnostic tool to monitor the thermal transport property of molten salts flowing in the loop. It also opens the potential to apply the MPR technique to on-site facilities for *in-situ* measurement of other high-temperature HTFs.

**Acknowledgements**

This material is based upon work supported by the U.S. Department of Energy's Office of Energy Efficiency and Renewable Energy (EERE) under Solar Energy Technologies Office (SETO) Agreement Number DE-EE0008379. The views expressed herein do not necessarily represent the views of the U.S. Department of Energy or the United States Government.

**Figure and Table Lists**

**Table 1** The properties of the eutectic chloride salt NaCl-KCl-MgCl$_2$ (wt % 15.11–38.91–45.98). It is from reference (Wang et al., 2021)

| Properties | Correlations as a function of $T$ (°C) |
|---|---|
| Density $\rho_o$ (kg m$^{-3}$) | $1958.8438 - 0.56355 \times T$ |
| Specific heat capacity $c_{p,o}$ (J g$^{-1}$ K$^{-1}$) | $1.30138 - 0.005 \times T$ |
| Dynamic viscosity $\mu$ (Pa·s) | $7.0645 \times 10^{-4} \exp\left(\frac{1204.11348}{T+237.15}\right)$ |
| Thermal conductivity $k_o$ (W m$^{-1}$ K$^{-1}$) | $0.5822 - 2.6 \times 10^{-4} \times T$ |

**Table 2** Summary of the experimental conditions for MPR flowing molten NaCl-KCl-MgCl$_2$ measurements

| Temperature $T$ (°C) | Average flow velocity $u_b$ (mm s$^{-1}$) |
|---|---|
| 520 | 375 ± 42 |
|  | 658 ± 33 |
|  | 1000 ± 33 |
| 580 | 266 ± 28 |
|  | 448 ± 35 |
|  | 628 ± 38 |
|  | 971 ± 16 |
|  | 1040 ± 34 |

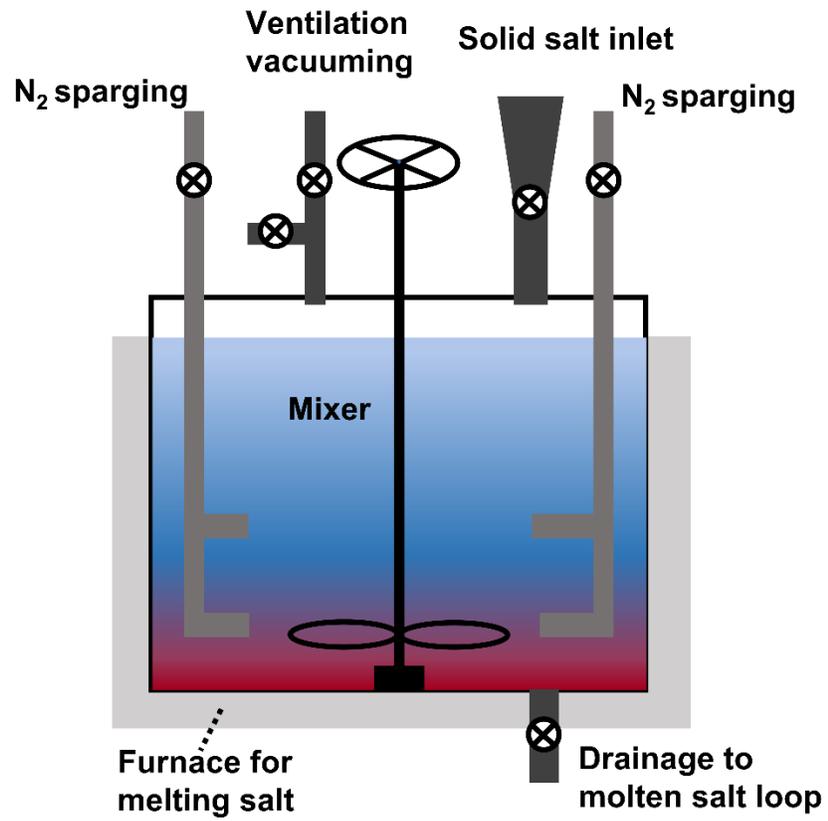

**Figure 1** Salt melting and purification in salt reservoir

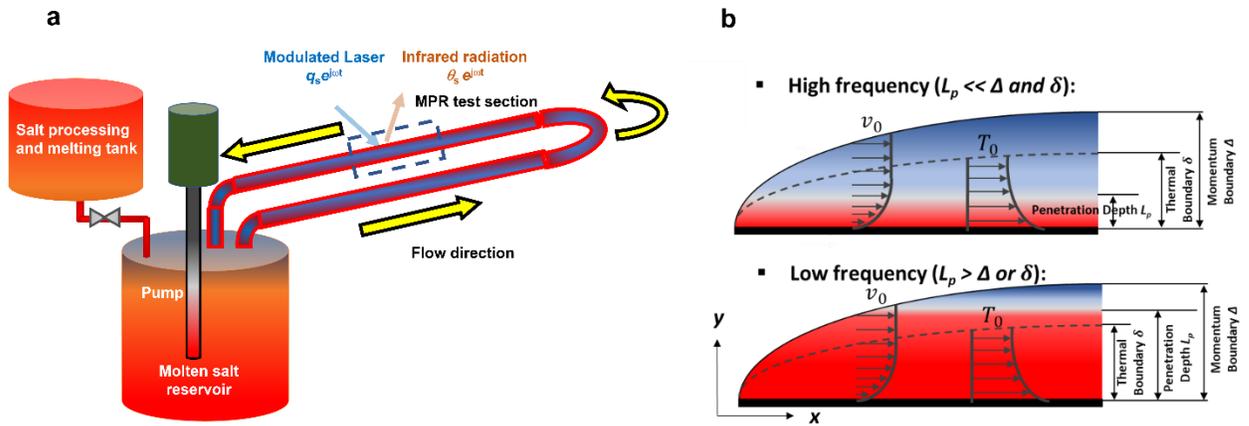

**Figure 2 (a)** Schematic of molten chloride salt loop and the MPR *in-stiu* measurement on the test section; and **(b)** Schematics of momentum boundary, thermal boundary and thermal penetration depth in the MPR measurement

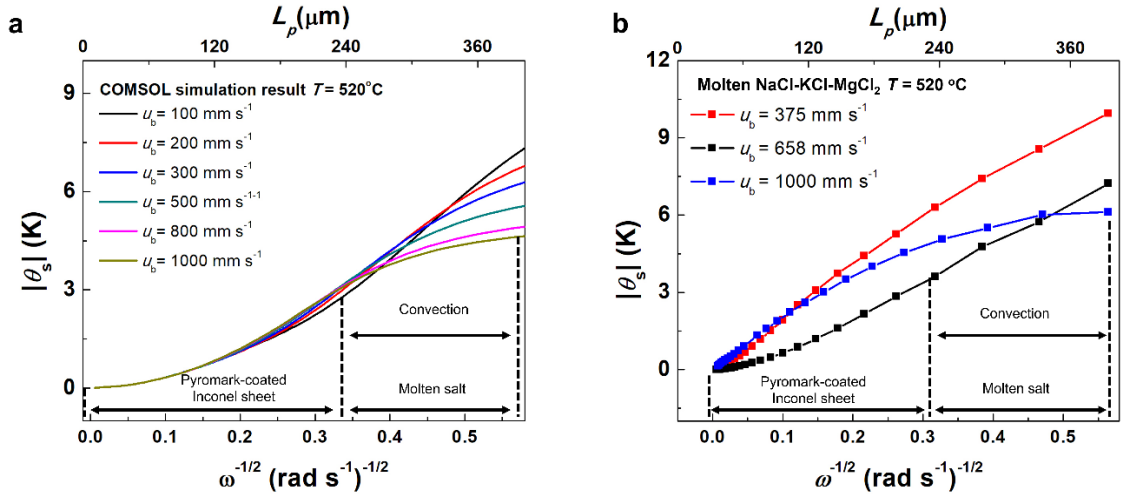

**Figure 3 (a)** COMSOL simulation result of $|\theta_s|$ *vs.* $\omega^{-\frac{1}{2}}$ for flowing molten NaCl-KCl-MgCl$_2$ measurement with various $u_b$ ranging from 100 - 1000 mm s$^{-1}$; **(b)** plot of $|\theta_s|$ versus $\omega^{-\frac{1}{2}}$ of flowing NaCl-KCl-MgCl$_2$ at $T$ = 520 °C with different $u_b$: 375, 658, and 1000 mm s$^{-1}$

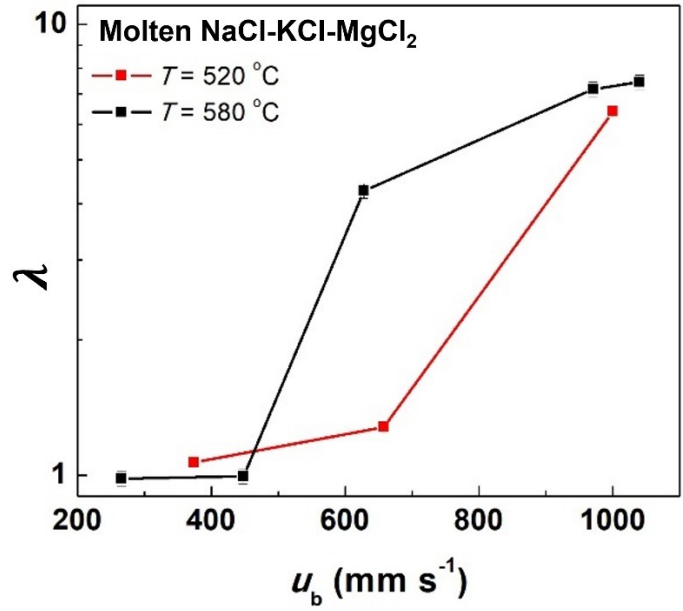

**Figure 4** Plot of the normalized thermal conductivity $\lambda = \frac{k_{eff}}{k_o}$ of flowing salt as a function of fluid flow velocity $u_b$.

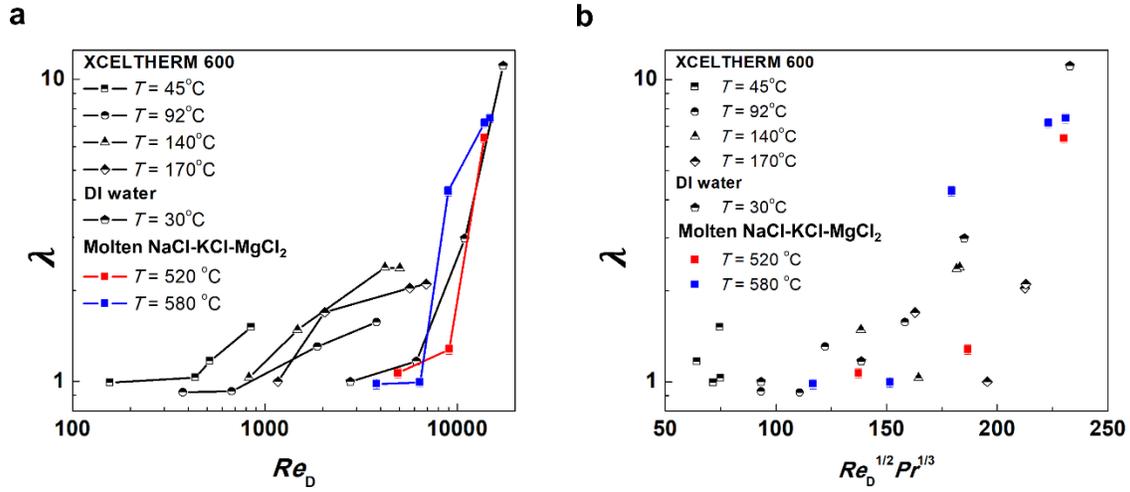

**Figure 5** Analysis of the forced convection effect in flowing molten salt. **(a)** normalized thermal conductivity $\lambda$ as a function of $Re_D$ number for molten salt; and **(b)** normalized thermal conductivity $\lambda$ of flowing molten salt as a function of $Re_D^{\frac{1}{2}} Pr^{-\frac{1}{3}}$ at different temperatures. Our previous flowing D.I. water and Xceltherm 600 measurement data using the MPR method are included for comparison (Zeng et al., 2021a).

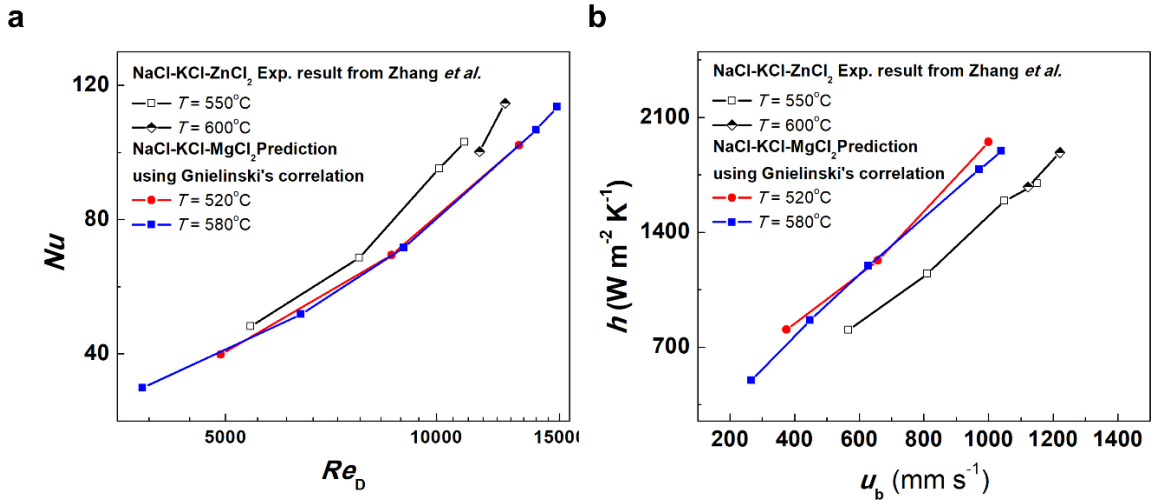

**Figure 6** The predicted **(a)** Nusselt number (*Nu*) and **(b)** Heat transfer coefficient (*h*) of molten NaCl-KCl-MgCl$_2$ using Gnielinski's correlation. The results from Zhang *et al*. (Zhang et al., 2022) of flowing molten NaCl-KCl-ZnCl$_2$ are plotted on the same graphs for comparison.

Supplementary Information

In-*situ* Thermophysical Measurement of Flowing Molten Chloride Salt Using Modulated Photothermal Radiometry

Ka Man Chung[1,#], Ye Zhang[3,#], Jian Zeng[2], Fouad Haddad[3], Sarath Adapa[2], Tianshi Feng[2], Peiwen Li[3], Renkun Chen[1,2,*]

[1]Material Science and Engineering Program, University of California, San Diego, La Jolla, California 92093, United States

[2]Department of Mechanical and Aerospace Engineering, University of California, San Diego, La Jolla, California 92093, United States

[3]Department of Aerospace and Mechanical Engineering, The University of Arizona, Tucson, AZ 85721, United States

\* Corresponding author.

Email address: rkchen@ucsd.edu (R. Chen)

# These authors contributed equally

**Table S1** Specifications of instruments used in the high temperature molten salt loop

| Name | Measurable parameter | Accuracy | Model and maker | Range |
|---|---|---|---|---|
| Furnace for molten salt reservoir | Volume | N/A | Model KM714, Skutt Ceramic Products, Inc. | T <1000 °C |
| Pipe | Outer diameter & wall thickness | OD ±0.127 mm; wall tolerance ±0.051 mm | Hastelloy C276, American Special Metal. | 25.4 mm O.D. & 1.2446 mm thick. T <800 °C |
| Molten salt pump | Volume flow rate | N/A | Nagle Pumps, Inc. | 0-0.63 (L s$^{-1}$) T <1000 °C |
| Ultrasonic flow meter | Flow velocity | ±1.6 % of reading in addition to ±0.01 m s$^{-1}$ | FLUXUS ADM 7407, Flexim Americas Co. | 0.01-25 m s$^{-1}$, T < 650 °C |
| Heating tapes | Power | N/A | STH051-080, Omega Engineering, Inc. | 0-627 W, T < 760 °C |
| Ceramic Heaters | Power | N/A | VF-180-3-19.75-V-S, Thermcraft Inc. | 0-1615 W, T < 1200 °C |
| Micro Fusion SCR Power Controller | Power | ±0.5 % of reading | uF1HXTA0-16-P1R00, Control Concepts Inc. | 0-10 kW |
| Digital temperature controllers | Temperature | ±3 °C of reading | SDC240KC-A, by Semi-conductor Smart Solutions, LLC. | T < 700 °C |
| Thermocouples | Temperature | ±1.1 °C or 0.4 % of reading | KQXL-116G-24, KQXL-116G-18, Omega Engineering, Inc. | 0-1038 °C |
| Multifunction I/O Device | Voltage | ± 0.03 % of reading | NI cDAQ-9178, NI9229, National Instruments. | -60 V to +60 V |

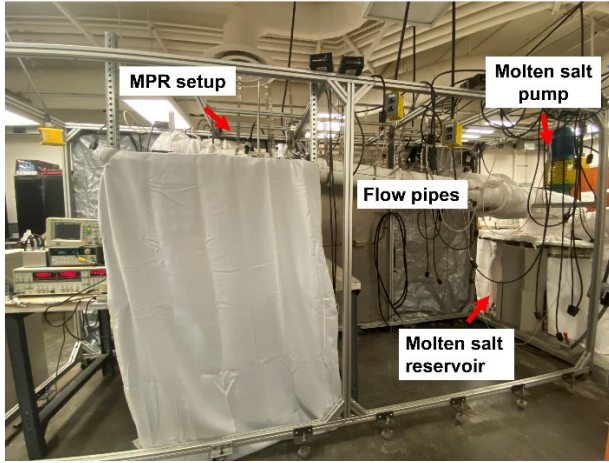 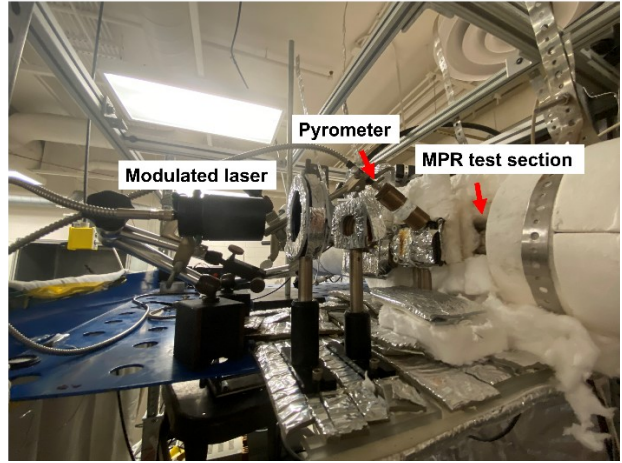

**Figure S1 (a)** Overview of the molten salt loop; and **(b)** close-up of the MPR setup that is integrated to the molten salt flow loop

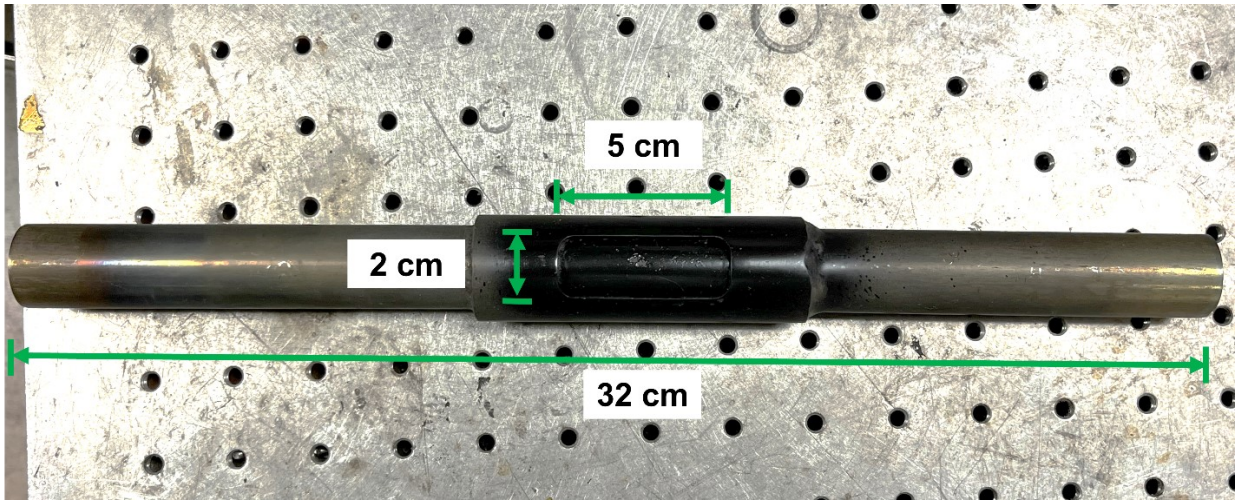

**Figure S2** Photograph of the MPR test section

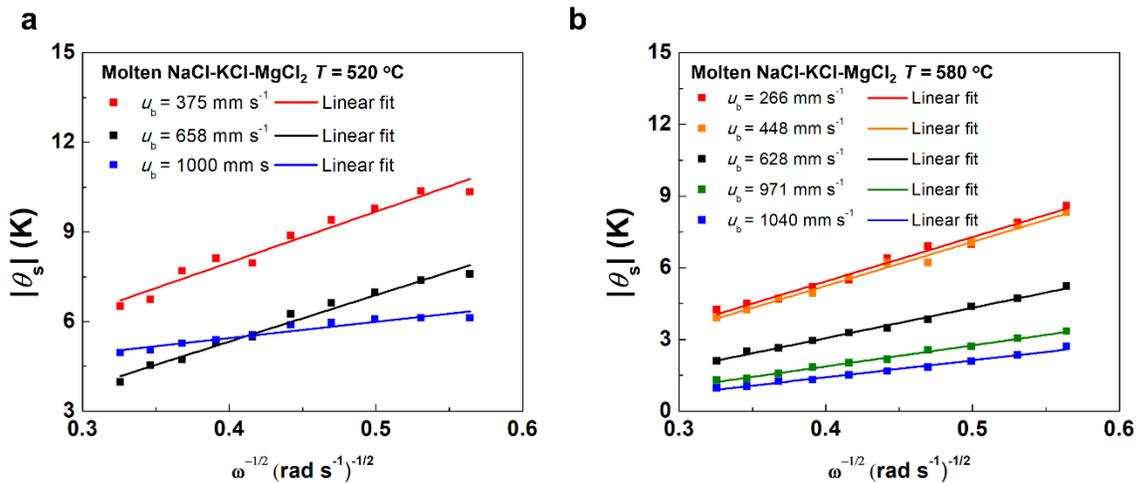

**Figure S3** Plot of $|\theta_s|$ vs. $\omega^{-\frac{1}{2}}$ of flowing NaCl-KCl-MgCl$_2$ **(a)** at $T = 520$ °C with different $u_b$ ranging from 375 to 1000 mm s$^{-1}$; **(b)** at $T = 580$ °C with different $u_b$ ranging from 266 to 1040 mm s$^{-1}$

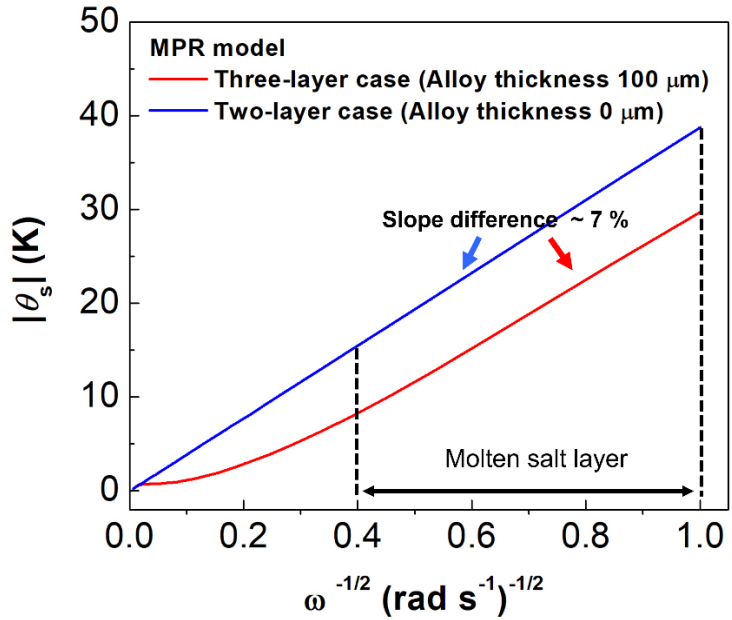

**Figure S4** $|\theta_s|$ vs. $\omega^{-\frac{1}{2}}$ plots of a three-layer structure (Pyromark coating, 100 μm Inconel alloy, and molten salt layer) and a two-layer structure (Pyromark coating, and molten salt layer) simulated by the MPR model. The input parameters are the thermophysical properties of Pyromark coating, Inconel alloy and molten salts at $T$ = 500 °C (Wang et al., 2021; Zeng et al., 2021b). The slopes of the curves in the low-frequency region show a deviation of around 7%.